\documentclass[12pt,preprint]{aastex}

\begin{document}

\title{Probing the faint end of the quasar luminosity function at $z \sim$ 4 in the COSMOS field}
\author{H. Ikeda,\altaffilmark{1} T. Nagao,\altaffilmark{1,2} K. Matsuoka,\altaffilmark{1,3} Y. Taniguchi,\altaffilmark{2} Y. Shioya,\altaffilmark{2} J. R. Trump,\altaffilmark{4} P. Capak,\altaffilmark{5,6} A. Comastri,\altaffilmark{7} M. Enoki,\altaffilmark{8} Y. Ideue,\altaffilmark{1,3} Y. Kakazu,\altaffilmark{5} A. M. Koekemoer,\altaffilmark{9} T. Morokuma,\altaffilmark{3,10} T. Murayama,\altaffilmark{11} T. Saito,\altaffilmark{12} M. Salvato,\altaffilmark{13} E. Schinnerer,\altaffilmark{14} N. Z. Scoville,\altaffilmark{5} and J. D. Silverman\altaffilmark{12} }

\altaffiltext{1}{Graduate School of Science and Engineering, Ehime University, Bunkyo-cho, Matsuyama 790-8577, Japan; email: ikeda@cosmos.phys.sci.ehime-u.ac.jp}
\altaffiltext{2}{Research Center for Space and Cosmic Evolution, Ehime University, Bunkyo-cho, Matsuyama 790-8577, Japan}
\altaffiltext{3}{Research Fellow of the Japan Society for the Promotion of Science}
\altaffiltext{4}{UCO/Lick Observatory, Department of Astronomy and Astrophysics, University of California, Santa Cruz, CA 95064}
\altaffiltext{5}{California Institute of Technology, MC 105-24, 1200 East California Boulevard, Pasadena, CA 91125}  
\altaffiltext{6}{Spitzer Science Center, 314-6 Caltech, 1201 East California Boulevard, Pasadena, CA 91125}
\altaffiltext{7}{INAF-Osservatorio Astronomico di Bologna, via Ranzani 1, 40127 Bologna, Italy} 
\altaffiltext{8}{Faculty of Bussiness Administration, Tokyo Keizai University, 1-7-34 Minami-cho, Kokubunji, Tokyo 185-8502, Japan}
   \altaffiltext{9}{Space Telescope Science Institute, 3700 San Martin Drive, Baltimore, MD 21218}
 \altaffiltext{10}{Institute of Astronomy, Graduate School of Science, University of Tokyo, 2-21-1 Osawa, Mitaka 181-0015, Japan}
 \altaffiltext{11}{Astronomical Institute, Graduate School of Science, Tohoku University, Aramaki, Aoba, Sendai 980-8578, Japan}
 \altaffiltext{12}{Institute for the Physics and Mathematics of the Universe (IPMU), The University of Tokyo, Kashiwa 277-8582, Japan}
 \altaffiltext{13}{Max-Planck-Institut f\"ur Plasmaphysik, Boltzmanstrasse 2, D-85741 Garching, Germany}
\altaffiltext{14}{Max-Planck-Institut f\"ur Astronomie, K\"onigstuhl 17, D-69117 Heidelberg, Germany}

\shortauthors{Ikeda et al.}    

\begin{abstract}
 We searched for quasars that are $\sim$ 3 mag fainter than the SDSS quasars in the redshift range 3.7 $\lesssim$ {\it z} $\lesssim$ 4.7 in the COSMOS field to constrain the faint end of the quasar luminosity function. Using optical photometric data, we selected 31 quasar candidates with 22 $\textless$ $i\arcmin$ $\textless$ 24 at {\it z} $\sim$\ 4. We obtained optical spectra for most of these candidates using FOCAS on the Subaru telescope, and identified 8 low-luminosity quasars at $z\sim4$. In order to derive the quasar luminosity function (QLF) based on our spectroscopic follow-up campaign, we estimated the photometric completeness of our quasar survey through detailed Monte Carlo simulations. Our QLF at $z\sim4$ has a much shallower faint-end slope ($\beta=-1.67^{+0.11}_{-0.17}$) than that obtained by other recent surveys in the same redshift. Our result is consistent with the scenario of downsizing evolution of active galactic nuclei inferred by recent optical and X-ray quasar surveys at lower redshifts.
 \end{abstract}

\newpage

\section{Introduction}
The quasar luminosity function (QLF) is one of the most important tools to constrain the evolution of supermassive black holes (SMBHs). A number of luminous quasars have been found up to $z\sim6$ by the Sloan Digital Sky Survey (SDSS; Fan et al. 2000, 2001, 2003, 2004, 2006; Richards et al. 2006; Goto 2006; Jiang et al. 2008, 2009) and 2dF quasar surveys (Boyle et al. 2000; Croom et al. 2004). However, low-luminosity quasars\footnote{Although quasars have been defined as relatively high-luminosity AGNs with $M_{\rm B} <$ |21.5 + 5 log $h_0$ historically (Schmidt \& Green 1983), we do not adopt such a solid criterion to distinguish these two populations.} at $z > 4$ have not yet been well studied. In contrast, the QLF at $z\lesssim3$ is well quantified in a wide luminosity range (e.g., Croom et al. 2009) and is best represented by a double power law (e.g., Boyle et al. 1988; Pei 1995). Accordingly the bright-end slope $\alpha$ has been already well measured; the typical measured value is around $-3.1$ at $z<2.4$, and flattens to $\alpha \gtrsim-2.37$ by $z = 5$ (Richards et al. 2006). More interestingly, recent studies on the optical QLF show that the activity in low-luminosity active galactic nuclei (AGNs) peaks at a lower redshift than that of more luminous AGNs (Croom et al. 2009). By assuming that the brighter AGNs have the more massive SMBHs, this process can be interpreted as AGN  (or  SMBH) downsizing. The AGN downsizing is also reported by X-ray surveys (Ueda et al. 2003; Hasinger et al. 2005; see also Brusa et al. 2009). However, the lack of low-luminosity quasars at $z>4$ leaves the faint-end slope of the $z>4$ QLF very poorly constrained. Consequently it is not understood how low-luminosity quasars evolve at high redshifts, or if downsizing is also present in the earlier universe. 

Motivated by these issues, we have searched for low luminosity quasars at $z\sim4$ in the COSMOS field (Scoville et al. 2007). Throughout this paper we use a $\Lambda$ cosmology with $\Omega_m$ = 0.3,\  $\Omega_{\Lambda}$ = 0.7, and the Hubble constant of $H_0$ = 70 km s$^{-1}$ Mpc$^{-1}$.\\

\section{The Sample}
\subsection{The Data}
We select the quasar candidates  by using color and morphology criteria. We use the official COSMOS photometric redshift catalog (Ilbert et al. 2009). This catalog covers an area of $\sim$ 2 $\rm deg^2$ and contains several photometric magnitudes including the total magnitude (${\tt MAG\_AUTO}$) of the $i'$-band and $3^{''}$ diameter aperture magnitudes of the CFHT $u^*$-band as well as Subaru Suprime-Cam $g'$-, $r'$-, and $i'$-bands. Details of the Suprime-Cam observations and the COSMOS photometric catalog are given in Taniguchi et al. (2007) and in  Capak et al. (2007), respectively. The 5$\sigma$ limiting AB magnitudes are $u^*$ = 26.5, $g'$ = 26.5, $r'$ = 26.6, and $i'$ = 26.1. 
Since we also use the Advanced Camera for Surveys (ACS) catalog  (Koekemoer et al. 2007; Leauthaud et al. 2007) to separate galaxies from point sources, our survey area is restricted to the area mapped with ACS ($1.64\ \rm deg^2$).

\subsection{Quasar Candidate Selection}
A quasar at $z\sim4$ shows a Lyman break in its spectral energy distribution (SED) that falls between the wavelengths of  the $g'$ and $r'$ filters, making a $g'-r'$ color redder than their $r'-i'$ color. We utilize this characteristic to select candidates of low-luminosity quasars. Here the typical quasar colors as a function of redshift are necessary to define the quasar color-selection criteria. Therefore,
we created model quasar spectra following the procedure generally adopted (e.g., Fan 1999; Richards et al. 2006), and derived the $u^*-g'$, $g'-r'$, and $r'-i'$ colors of the model quasars at  redshifts between $0<z<6$. For the model quasar spectra we assume typical values of the power-law slope $\alpha_{\nu}$ ($f_{\nu}\propto\nu^{-\alpha_{\nu}}$) and Ly$\alpha$ rest-frame equivalent width (EW) to be 0.46 and 90\AA, respectively (Vanden Berk et al. 2001).
 In Fig. 1, the simulated colors of the model quasars are shown in the $r'-i'$ versus $g'-r'$ diagram. We select our candidates of quasars using the following criteria: (1) $22<i' {\rm ({\tt MAG\_AUTO})}<24$, (2) $r'-i'<0.42(g'-r')-0.22$, (3) $u^*-g' \geq 2.0$, and (4) $g'-r'>1.0$. where we use {\tt MAG\_AUTO} measured by SExtractor (Bertin \& Arnouts 1996) in eq. (1) instead of $3^{\prime\prime}$ aperture magnitude, because the target luminosity should be calculated from total magnitudes, not from aperture magnitudes.  The criterion (2) is used to select quasars efficiently without significant contamination of stars. To remove possible foreground contaminations further, we introduce the additional criteria (3) and (4). These latter two color thresholds (i.e., criteria (3) and (4)) are determined by taking empirical color distributions of quasars at $z\sim4$ into account (Richards et al. 2006). We also exclude spatially-extended objects defined by Leauthaud et al. (2007). As a result, we obtained 31 quasar candidates among 7318 point sources with $22 < i'(\tt MAG\_AUTO) < \rm 24$.

\section{Spectroscopic Observation}
The spectroscopic follow-up observations of the quasar candidates have been carried out at the Subaru telescope with FOCAS (Kashikawa et al. 2002) on 7--11 January 2010. We used the 300 grating with the SO58 filter, whose wavelength coverage is $\rm 5800\AA$ $\textless$ $\rm \lambda_{obs}$  $\textless$ $\rm10000\AA$ ($\rm1040\AA$ $\lesssim$ $\rm \lambda_{rest}$ $\lesssim$ $\rm1590\AA$ at $z=4$). We used a $0 \farcs 8$-width slit, resulting in a wavelength resolution of $R\sim700$ ($\Delta v$ $\sim430$ \rm km $\rm s^{-1}$) as measured by night sky emission lines. The typical seeing was $\sim0 \farcs 7$. We observed 28 of the 31 candidates, and obtained useful spectra for 23 of them. The remaining 5 objects show no signal, possibly due to insufficient exposure time (the typical exposure time of these objects is 1800 seconds). 

Standard data reduction procedures were performed using IRAF. After sky subtraction,
we extracted one-dimensional spectra by adopting an aperture size of $1 \farcs 8$. The spectra of 23 objects were flux-calibrated using spectrophotometric standard stars. We found that eight show strong and broad Ly$\alpha$ and {C {\sc iv}  emission lines, suggesting that these eight objects are indeed quasars at $z\sim4$. Of the remaining 15 objects, one spectrum shows narrow {C {\sc iv}}, {He {\sc ii}}, {O {\sc iii]}}, and {C {\sc iii]}} emission lines (type-II quasar) at $z\sim3.5$. The spectra of three other objects show only narrow emission lines without any high-ionization lines such as {C {\sc iv} present, being consistent with Ly$\alpha$ emitting galaxies at $z\sim4$. The remaining 11 objects appear to be consistent with Galactic late-type stars (from K4 to M0 stars). The photometric properties and the measured redshifts of the identified quasars are summarized in Table 1. An example of the reduced spectra of the identified quasars is shown in Fig. 2.

\section{Quasar Luminosity Function}
Our spectroscopic run found eight quasars at $z\sim4$. Accounting for the possibility that some of the photometric candidates without spectra could be also quasars (based on the success rate of our spectroscopic run), the \textquotedblleft corrected\textquotedblright \ number of the quasars in our survey are $(4.35\pm2.01)$ $\times$ 10$^{-7}$ Mpc$^{-3}$ mag$^{-1}$ for $-24.09 < M_{1450} < -23.09$ and $(7.32\pm2.97)$ $\times$ 10$^{-7}$ Mpc$^{-3}$ mag$^{-1}$ for $-23.09 < M_{1450} < -22.09$, respectively. To derive the QLF based on these results, we need to compute the effective comoving volume. The effective comoving volume of the survey is calculated as:\\
\addtocounter{equation}{4}
\begin{equation}
	V_{\rm eff} (m_{i'})= d \Omega \int_{z = 0}^{z = \infty}C(m_{i'}, z)\frac{dV}{dz}dz,   
	\end{equation}where $d\Omega$ is the solid angle of the survey and $C(m_{i'}, z)$ is the photometric completeness. Here the photometric completeness is calculated by modeling quasar spectra as described in Section 2.2, but taking also the intrinsic variation of the continuum slope and the equivalent width (EW) of emission lines into account. We assume a Gaussian distribution of the power-law slope $\alpha_{\nu}$ ($f_{\nu}\propto\nu^{-\alpha_{\nu}}$) and Ly$\alpha$ EWs, with means of 0.46 and 90{\AA} (the same as those in Section 2.2), and a standard deviation of 0.30 and 20{\AA}, respectively (Vanden Berk et al. 2001). We created 1000 quasar spectra at each $\Delta$$z$ = 0.01 for the redshift range $0<z<6$. The effects of intergalactic absorption by neutral hydrogen were corrected by adopting the extinction model of Madau (1995). Then, we calculated the colors of the model quasars in the observed frame. We put the simulated quasar images into Subaru FITS images and measure their colors. Applying the color-selection criteria (1) -- (4), we infer the photometric completeness. The redshift range where the completeness is moderately good is 3.7 $\lesssim$ {\it z} $\lesssim$ 4.7. Specifically, the inferred completeness is $\sim 0.7$ for quasars with $i' = 22.5$, and $\sim0.5$ for those with $i' = 23.5$.	
Given the effective comoving volume, the comoving space density and its standard deviation are calculated through the standard procedure (see, e.g., Fan et al. 2001). The obtained QLF is plotted in Fig. 3. Although the redshift range of SDSS data (4.0 $\leq$  ${\it z}$ $\leq$ 4.5; Richards et al. 2006) is slightly different from our study (3.7 $\lesssim$ {\it z} $\lesssim$ 4.7), 
we apply the weighted least-squares fit to the space density of quasars at $z\sim4$ inferred by our study and SDSS, adopting the following double power-law function:
\begin{equation}
\Phi(M_{1450}, z) = \frac{\Phi(M^*_{1450})}{10^{0.4(\alpha+1)(M_{1450}-M^*_{1450})}+10^{0.4(\beta+1)(M_{1450}-M^*_{1450})}},
\end{equation}
where $\alpha$, $\beta$, $\Phi^*{(M_{1450})}$, and $M^*_{1450}$ are the bright-end slope, the faint-end slope, the normalization of the luminosity function, and the characteristic absolute magnitude, respectively. Among the four parameters, the bright-end slope ($\alpha$) is fixed to be $\alpha = -2.58$ based on the $z\sim4.2$ SDSS results (Richards et al. 2006). The best-fit parameters are $\Phi$($M^*_{1450}$) = $(3.20 \pm \rm0.24)$ $\times$ 10$^{-7}$ Mpc$^{-3}$ mag$^{-1}$, $M^*_{1450} = -24.40 \pm \rm 0.06$, and $\beta = -1.67^{+0.11}_{-0.17}$. The fitting result is shown in Fig. 3.  

Here we compare our QLF at $z\sim4$ with another QLF for the same redshift recently reported by Glikman et al. (2010). They searched for quasars at $z\sim4$ using the data of the NOAO Deep Wide-Field Survey (NDWFS; Jannuzi \& Dey 1999) and the Deep Lens Survey (DLS; Wittman et al. 2002), whose total area is $\rm 3.76 \ deg^2$. By combining their own data with the SDSS results, they derived a faint-end slope of $\beta = -2.3\pm0.2$. This is significantly steeper than our result ($\beta = -1.67^{+0.11}_{-0.17}$). This discrepancy could be possibly caused by the fact that only six spectra have been obtained for a total of 117 photometric quasar candidates at $R>23$ in the sample of Glikman et al. (2010) and there is a possibility that a large fraction of their photometric quasar candidates at $R>23$ are contaminants (Ly$\alpha$ emitting galaxies and Lyman break galaxies), since such high-$z$ star-forming galaxies could be also selected through the Lyman-break selection criteria. While we removed such star-forming galaxies by using their ACS images, Glikman et al. (2010) used ground-based images to distinguish point sources and extended sources. Note that the number density of quasars at $R>23$ in Glikman et al. (2010) is roughly consistent with that of Lyman break galaxies at $z\sim4$ with similar luminosity (e.g., Ouchi et al. 2004). The fact that the two luminosity functions are roughly consistent suggest a large contaminant fraction in their quasar candidates. It should be also mentioned that Cristiani et al. (2004) reported a quasar number count at $z\sim 3-5$, which requires a flattened faint-end slope in the QLF. 

 We now discuss the evolution of the quasar space density. To compare the quasar space density with previous studies, we need to estimate the quasar space density in the same magnitude bins. Therefore we re-calculated the space density of quasars with $-23.5<M_{\rm 1450}<-22.5$. Fig. 4 shows the redshift evolution of the quasar space density for different absolute magnitude bins. Although there are a number of low-luminosity quasar surveys at  $z\sim3$ (Wolf et al. 2003; Hunt et al. 2004; Fontanot et al. 2007; Bongiorno et al. 2007), we plot only the results of the 2dF-SDSS LRG and Quasar Survey (2SLAQ; Croom et al. 2009), the Spitzer Wide-area Infrared Extragalactic Legacy Survey (SWIRE; Siana et al. 2008), and SDSS (Richards et al. 2006), in order to avoid data with large statistical errors. Although most studies at $z < 3$ have suggested consistent results (AGN downsizing), the situation becomes rather controversial at $z > 3$. Specifically, the results of this study and by Glikman et al. (2010) show completely different pictures. If the result of Glikman et al. (2010) is correct, the high number density of low-luminosity quasars (that may have less-massive SMBHs) at $z\sim4$ would correspond to the \textquotedblleft seeds\textquotedblright of luminous quasars (with high-mass SMBHs) at $z\sim2-3$. However, our results do not show any evidence of such a break down of the downsizing scenario at $z\sim4$, suggesting that numerous seeds of high mass SMBHs should exist at even higher redshift ($z > 5$). Note that the type-II (i.e., obscured AGN) fraction is reported to be higher in lower-luminosity samples at higher redshifts (La Franca et al. 2005). Therefore one may suspect whether the possible abundant obscured population affects the analysis of the (un-obscured) quasar number density evolution, since we are now focusing on high redshift low-luminosity AGNs. However, given the redshift and luminosity of our sample, the inferred obscuration fraction is $\sim0.3$ at most (Hasinger 2008). This suggests that our analysis is not significantly affected by the obscured population. 
 To obtain a more conclusive understanding of the quasar evolution in the early universe, extensive surveys of high-$z$ low-luminosity quasars are definitely required. The next-generation wide-field prime-focus camera for the Subaru Telescope (Hyper Suprime-Cam; Miyazaki et al. 2006) and Large Synoptic Survey Telescope (LSST; Ivezi\'c et al. 2008) will provide us with such opportunities in very near future.

\acknowledgments
We would like to thank the Subaru staff for their invaluable help and all members of the COSMOS team. TM , KM , and YI are financially supported by the Japan Society for the Promotion of Science (JSPS) through the JSPS Research Fellowship.



\begin{table}[h]

\renewcommand{\tabcolsep}{1pt}

\caption{Redshift and optical photometry of the eight spectroscopically confirmed $z\sim4$ quasars}
\begin{tabular}{ccc@{\hspace{0.5cm}}c@{\hspace{0.5cm}}c@{\hspace{0.5cm}}c@{\hspace{0.5cm}}c@{\hspace{0.5cm}}c@{\hspace{0.5cm}}c@{\hspace{0.5cm}}c@{\hspace{0.5cm}}c} \hline
       &  & &Redshift&  $u^{*a}$ & $g'$ & $r'$ & $i'$ &$i'$ ({\tt MAG\_AUTO}) &$\rm Exp. Time^{\it b}$ \\
       Number &&& $z_{\rm spec}$$^{\it c}$ &(mag) & (mag)& (mag) & (mag)& (mag)&(sec) \\ \hline
        1 & && 3.89 & \textgreater27.49 & 25.05 & 23.56 & 23.51 &23.35 &  6000     \\
        2 & && 4.14 & \textgreater27.49 & 24.75 & 23.10 & 22.79  &22.55 &    6000      \\
        3 & && 3.56 & \textgreater27.49 & 25.82 & 24.42 & 24.09  &23.91 & 3600\\ 
        4 & && 4.20 & \textgreater27.49 & 27.27 & 24.09 & 23.68 & 23.45&1800\\
        5 &  &&3.86 &   27.42 & 24.62 & 23.45 & 23.21 &23.04 &2400\\
        6 &  &&3.65 & 27.00   & 24.14 & 22.54 & 22.35 &22.16 &3600\\
        7 &  &&4.45 & \textgreater27.49 & 25.23 & 22.79 & 22.22 &  22.03&3600\\
        8 &  &&4.16 & \textgreater27.49 & 24.94 & 23.27 & 22.95 & 22.78&7200\\ \hline

        \end{tabular}\flushleft{$^{a}$ When the $u^*$-band magnitude is fainter than 2$\sigma$ limiting magnitude (= 27.49), the 2$\sigma$ lower limit is given.\\
        $^{b}$ Total on-source exposure time in the FOCAS spectroscopic observation.\\
        $^{c}$ $z_{\rm spec}$ is based on the {C {\sc iv} emission line.}}
\end{table}


\begin{figure}[h]
\begin{center}
\includegraphics[bb=-50 0 450 340,clip,width=15cm]{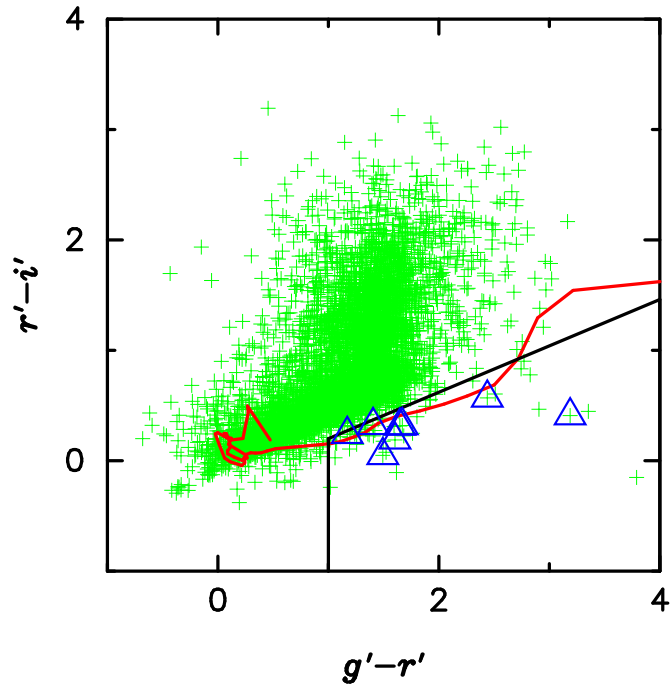}

\end{center}
\caption{Two color diagram between $r'-i'$ and $g'-r'$ used for our quasar selection. Green points are point sources with $22<i'({\tt MAG\_AUTO})<24$. Blue triangles (8 objects) are spectroscopically confirmed $z\sim4$ quasars. The  red line is the median track of modeled quasar colors as a function of redshift. The black solid line shows our photometric criteria used to select quasar candidates.}

\label{fig:Two color diagram used in our quasar selection.}

\end{figure}

\begin{figure}[h]
\begin{center}
\includegraphics[bb=0 50 650 800,clip,width=16cm,angle=270 ]{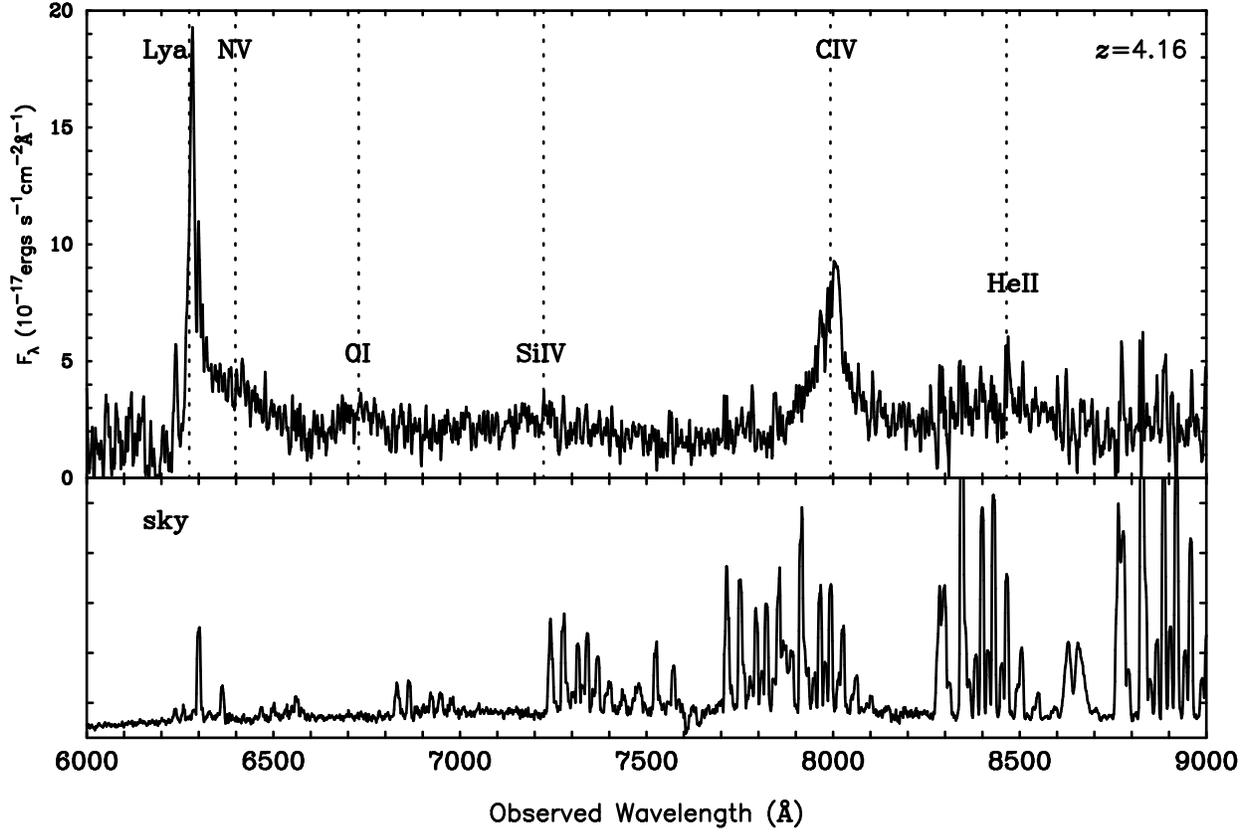}

\end{center}
\caption{Reduced spectrum of the quasar No. 8 (upper panel) and typical sky spectrum (lower panel). Dotted lines show the expected wavelengths of quasar emission lines: Ly$\alpha$ $\lambda1216$, {N {\sc v} $\lambda1240$}, {O {\sc i} $\lambda1304$}, {Si {\sc iv} $\lambda1400$}, {C {\sc iv} $\lambda1549$}, and {He {\sc ii} $\lambda1640$}.}

\label{ quasar spectrum}

\end{figure}

\begin{figure}[h]
\begin{center}
\includegraphics[bb=0 0 600 550,clip,width=12cm]{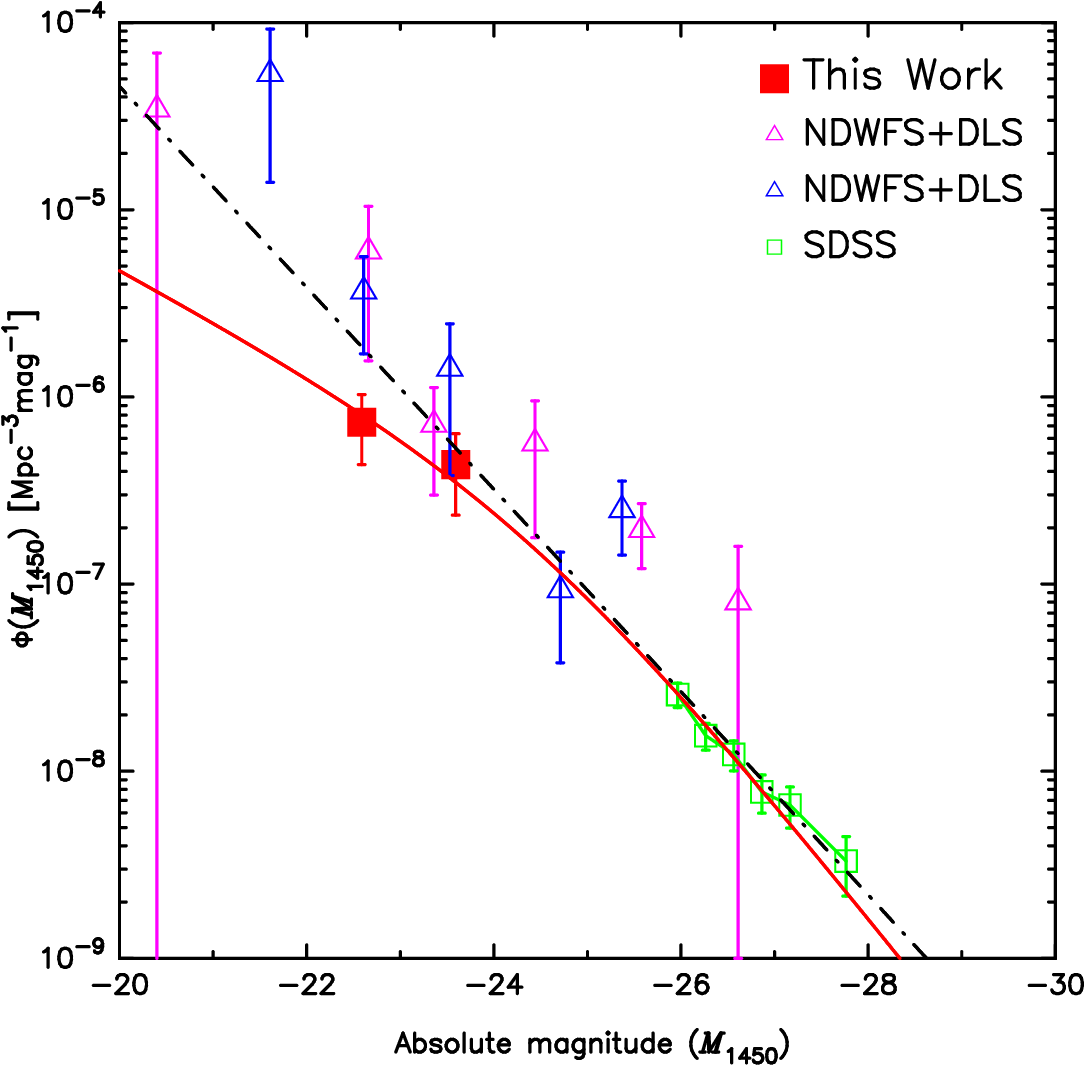}

\label{fig:quasar luminosity function}
\caption{The $z\sim4$ quasar luminosity function. The red squares show our results. Triangles are the data of NDWFS+DLS using  $M_{1450}$ from K-corrected magnitude (blue) and directly from the quasar spectra (magenta) (Glikman et al. 2010). Green squares are the SDSS results for the $z = 4.25$ bin (Richards et al. 2006). The red line shows the fit to our results (i.e., COSMOS) and the SDSS points. The black dash-dotted line shows the fit to the data from the NDWFS+DLS plus SDSS study (Glikman et al. 2010).}

\end{center}

\end{figure}

\begin{figure}[h]
\begin{center}
\includegraphics[bb=0 0 600 600,clip,width=12cm]{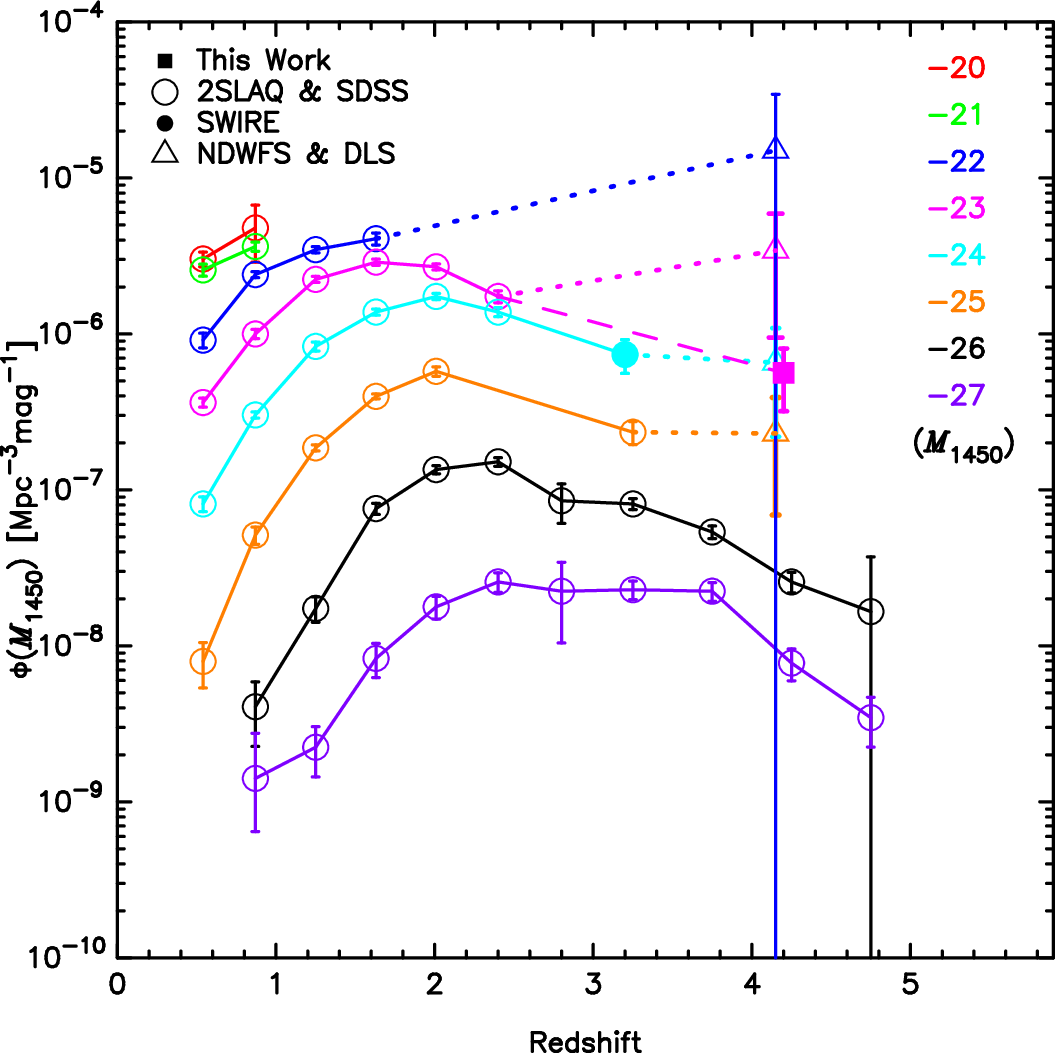}

\end{center}
\caption{The redshift evolution of the quasar space density. Red, green, blue, magenta, light blue, orange, black, purple lines are $M_{1450} = -20, -21, -22, -23, -24, -25, -26$, \rm and $-27$, respectively. Dotted lines show the combined 2SLAQ, SDSS, SWIRE, NDWFS and DLS QLF. Dashed line shows the combined this work (i.e., COSMOS) and 2SLAQ QLF.}

\label{fig:Redshift evolution of the quasar space density}

\end{figure}

\end{document}